# High frequency poroelastic waves in hydrogels

Piero Chiarelli and Claudio Domenici

*National Council of Research of Italy, Moruzzi, 1 - 56100 Pisa – Italy*

Antonio Lanatà and Marina Carbone

*Interdepartmental Research Center "E. Piaggio", Faculty of Engineering, University of Pisa,*

*via Diotisalvi, 2 -56126 Pisa, Italy*

**ABSTRACT**

In this work a continuum model for high frequency poroelastic longitudinal waves in hydrogels is presented.

A viscoelastic force describing the interaction between the polymer network and the bounded water present in such materials is introduced.

The model is tested by means of ultrasound wave speed and attenuation measurements in polyvinylalcohol hydrogel samples.

The theory and experiments show that ultrasound attenuation decreases linearly with the increase of the water volume fraction "$\beta$" of the hydrogel.

The introduction of the viscoelastic force between the bounded water and the polymer network leads to a bi-phasic theory showing an ultrasonic fast wave attenuation that can vary as a function of the frequency with a non-integer exponent in agreement with the experimental data in literature.

When $\beta$ tends to 1 (100% of interstitial water) due to the presence of bounded water in the hydrogel, the ultrasound phase velocity acquires higher value than that of pure water. The ultrasound speed gap at $\beta = 1$ is confirmed by the experimental results that show that it increases in less cross-linked gel samples that own a higher concentration of bounded water.

PACS numbers: 43.20.Jr, 43.20.Bi, 43.80.Cs, 43.80.Ev, 43.80.Qf

## I. INTRODUCTION

The present work is motivated in defining a reliable model for ultrasound (US) wave propagation both in hydrogels[1-5] as well as in extra cellular matrices of natural soft tissues[6].

The model can be a theoretical tool for the study and the characterization of tissues-mimicking phantom for US thermal therapy[7] and for the development of non-invasive assessment technique of soft tissue stiffness[8].

Most of the bi-phasic models on the acoustic wave propagation at high frequencies recently proposed[9-12] are mainly oriented to the geology and engineering.

As far as the US propagation in natural hydrogels is concerned, this is usually modeled by means of the wave equation that holds for water since they are mainly composed by it[5]. Even if this approach is sufficiently satisfying, it does not completely explain the experimental behavior of US.

There are many discrepancies between US propagation in water and in natural hydrogels of soft tissues both for transverse and longitudinal acoustic waves[6].

As far as longitudinal waves are concerned, the frequency law[13] $\nu^{(1+\delta)}$ applied to US absorption in biological gels, where $\nu$ is the US frequency, gives $\delta$ that ranges between ¼ and ½, while it holds that $\delta = 1$ for water.

The fractional value of $\delta$ cannot be explained by means of the established bi-phasic theories that lead to a frequency dependence equal to that of water with $\delta = 1$[14-15].

In order to solve the theoretical leak, in the present paper, a viscoelastic matrix-fluid interaction specific for hydrogels is introduced.

If we look at the structure of the hydrogel we see that it is somewhere between a solid and a liquid. It consists of polymers, or long chain molecules, which are cross-linked to create an entangled network immersed in a liquid medium that fills the intra-molecular interstices.

The properties of gels are strongly influenced by the interaction of these two components.

Firstly, gels were conceptualized as porous media consisting of two interpenetrating macroscopic substances (an elastic and porous solid matrix and a fluid). An elegant and satisfying theory based on this approach is the poroelastic model developed by Biot[14-17].

One of the main characteristics of the Biot theory is that the dissipative relaxation is mainly ascribed to the relative motion of fluid against solid, while dissipations within fluid and solid are neglected.

If at low frequencies, the friction of the fluid that moves against the solid network leads to a "diffusional" wave that gives theoretical forecasts that well agree with many experimental results[18-23], such a solid-fluid interaction scheme becomes inadequate at high frequencies.

In order to obtain a more satisfying model to describe how ultrasounds propagate in hydrogels, we propose a new mechanism of interaction between the fluid (water) and the solid matrix (polymer network).

The main mathematical statement, introduced here, takes into account the presence of the bounded water attached to the polymeric network of the hydrogel. The fluid-matrix interaction is modeled by ascribing a viscoelastic force between the polymer matrix and the bounded water around it while a viscous force between the "polymer-bounded water aggregate" and the interstitial free water is assumed.

In the following we derive the phase velocity of longitudinal US plane waves and their attenuation in hydrogels (section II) by using the proposed fluid-matrix interaction.

Then, in section III, we compare the theoretical forecasts with the outcomes of the experimental measurements.

Since the percentile content of water in gels is often very high (up to 99%) the poro-elastic theories for hydrogels make historically use of the dilute matrix approximation[24-27] (i.e., $\beta \approx 1$). For sake of clarity, we maintain here this approximation taking care to check its range of applicability at the end.

## II. POROELASTIC WAVES IN HYDROGELS

When the viscous skin depth $(\eta_d / \pi \nu \rho_f)^{1/2}$ becomes smaller than the pore size as the frequency $\nu$ increases, the Biot's theory[14-15] takes the fluid-solid friction coefficient "$F_{(\nu)}$" into account to increase according to the law: $\lim_{\omega/\omega_c \to \infty} F_{(\nu)} \propto (\nu/\nu_c)^{1/2}$, where $\nu_c = f/\rho_f$, $f$ is zero frequency friction coefficient (i.e., the inverse of the hydraulic permeability of the mean)[25], $\rho_f$ is the fluid mass density and $\eta_d$ is its dynamic viscosity.

Given that pore dimensions are very small in gels, the crossover frequency $\nu_c$ is very high (for instance poly-vinyl alcohol-poly-acrylic acid hydrogels[26] show $\nu_c \approx 10^{13}$ Hz, close to the maximum frequency allowed in material media) and for usual US frequencies the factor $F_{(\omega)}$ in hydrogels should be always used in the low frequency limit (i.e., $F_{(\omega)} = 1$). This fact gives the US attenuation[15] proportional to $\nu^2$ while the experimental results follow the law[6] $\propto \nu^n$ with n ranging between 1,25 and 1,50.

In the frame of the bi-phasic approach one possible way out is to consider a different fluid-network interaction for hydrogels.

If we look at the hydrogel structure, it shows water molecules that are bounded to the matrix polymer chains by means of chemical interactions. Thence, a more appropriate matrix-fluid interaction scheme is assumed as follow:

i. a viscoelastic interaction between the bounded water and the polymer matrix (with an elastic constant $k$ and a friction coefficient $\eta$);

ii. a pure viscous interaction between the bounded water molecules (surrounding the polymer network) and the free water.

**A. Compressional waves for dilute poroelastic means**

Before introducing the specific hydrogel fluid matrix interaction, we derive the poroelastic longitudinal wave equations in the limit of very dilute matrix with incompressible solid and liquid constituents.

In this case the compressional elastic moduli of the fluid and the solid matrix can be assumed to be greater than all the other elastic moduli in the poroelastic longitudinal wave equations[14-17] that read

$$\nabla^2(P\varepsilon_{\alpha\alpha} + Qe_{\alpha\alpha}) = \partial^2(\rho_{11}\varepsilon_{\alpha\alpha} + \rho_{12}e_{\alpha\alpha})/\partial t^2 + \beta f F_{(\omega)} \partial(\varepsilon_{\alpha\alpha} - e_{\alpha\alpha})/\partial t \quad (1)$$

$$\nabla^2(Q\varepsilon_{\alpha\alpha} + Re_{\alpha\alpha}) = \partial^2(\rho_{12}\varepsilon_{\alpha\alpha} + \rho_{22}e_{\alpha\alpha})/\partial t^2 - \beta f F_{(\omega)} \partial(\varepsilon_{\alpha\alpha} - e_{\alpha\alpha})/\partial t \quad (2)$$

Where $\varepsilon_{ij}$ is the solid strain tensor and $e_{\alpha\alpha}$ is the trace of the liquid strain tensor; P, Q, and R are the poroelastic constants of the medium that can be measured by means of jacketed and unjacketed experiments[17]; $\beta$ is the pores volume fraction equating the fluid volume fraction and $\rho_{11}$, $\rho_{12}$ and $\rho_{22}$, are the mass density parameters defined as: $\rho_{11} + 2\rho_{12} + \rho_{22} = \rho$, $\rho_{11} + \rho_{12} = (1-\beta)\rho_s$, $\rho_{12} + \rho_{22} = \beta\rho_f$; where $\rho_s$ represents the solid mass density, while $\rho$ is the total mass density of the bi-phasic medium.

In the limit of high liquid phase content ($\beta \approx 1$) and incompressible constituents, the following conditions over the poroelastic constants and mass density parameters hold[14-16]

$$R \gg Q \gg P \tag{3}$$

$$Q/R \cong (1-\beta)/\beta \cong (1-\beta) \tag{4}$$

$$\rho_{12} = -\rho_{11} \ll \rho_{22} \tag{5}$$

By introducing the above approximations into the longitudinal wave equations we obtain

$$\nabla^2 Q e_{\alpha\alpha} = \partial^2 \rho_{11} (\varepsilon_{\alpha\alpha} - e_{\alpha\alpha})/\partial t^2 + \beta f F_{(\omega)} \partial(\varepsilon_{\alpha\alpha} - e_{\alpha\alpha})/\partial t \tag{6}$$

$$\nabla^2 (Q\varepsilon_{\alpha\alpha} + R e_{\alpha\alpha}) = \partial^2 (\rho_{12}\varepsilon_{\alpha\alpha} + \rho_{22} e_{\alpha\alpha})/\partial t^2 - \beta f F_{(\omega)} \partial(\varepsilon_{\alpha\alpha} - e_{\alpha\alpha})/\partial t \tag{7}$$

Given that the friction coefficient $f$ is very high[18, 20, 24-27] it follows that the displacement $|\varepsilon_{\alpha\alpha} - e_{\alpha\alpha}|$ of the "slow" wave (liquid and solid in counter-phase) is very small compared to that of the fast wave $|e_{\alpha\alpha}|$, so that it is possible to pose $|\varepsilon_{\alpha\alpha}| \approx |e_{\alpha\alpha}|$ and, hence

$$(\rho_{12}\varepsilon_{\alpha\alpha} + \rho_{22} e_{\alpha\alpha}) \cong (\rho_{22} + \rho_{12}) e_{\alpha\alpha} = \beta \rho_f e_{\alpha\alpha} \tag{8}$$

$$(Q\varepsilon_{\alpha\alpha} + R e_{\alpha\alpha}) \cong (R/\beta) e_{\alpha\alpha} \tag{9}$$

that introduced in (7) leads to

$$\nabla^2 \frac{R}{\beta} e_{\alpha\alpha} \cong \partial^2 (\beta \rho_f e_{\alpha\alpha})/\partial t^2 - \beta f F_{(\omega)} \partial(\varepsilon_{\alpha\alpha} - e_{\alpha\alpha})/\partial t \tag{10}$$

Assuming for dilute matrices the Barryman's condition[18]: $\rho_{11} \cong 0$, at zero order in $(1-\beta)$, (6,10) approximately read

$$\nabla^2 Q e_{\alpha\alpha} \cong \beta f F_{(\omega)} \partial(\varepsilon_{\alpha\alpha} - e_{\alpha\alpha})/\partial t \tag{11}$$

$$\nabla^2 \frac{R}{\beta} e_{\alpha\alpha} \cong \partial^2 (\beta \rho_f e_{\alpha\alpha})/\partial t^2 \tag{12}$$

Moreover, by introducing (12) in (11) the following relation between the "slow wave" and the fast wave displacements

$$F(\omega)\frac{\partial(\varepsilon_{\alpha\alpha} - e_{\alpha\alpha})}{\partial t} \cong \frac{1}{f}\frac{Q}{R}\frac{\partial^2(\beta\rho_f e_{\alpha\alpha})}{\partial t^2} \qquad (13)$$

is obtained.

Finally, by introducing (13) in (10), at first order in (1-β), we obtain

$$\nabla^2 \frac{R}{\beta} e_{\alpha\alpha} \cong \beta \,\partial^2 (\beta\rho_f e_{\alpha\alpha})/\partial t^2 \qquad (14)$$

that under the hypothesis of $\beta$ constant, reads:

$$\nabla^2 R e_{\alpha\alpha} \cong \beta^3 \,\partial^2 (\rho_f e_{\alpha\alpha})/\partial t^2 \qquad (15)$$

Eq. (15) represents a purely elastic wave of the first type that propagates in a medium with a speed *"c"*

$$c^2 = R/(\rho_f \beta^3) \approx c_f^2/\beta^3 \qquad (16)$$

where $c_f$ is the wave velocity in the fluid phase. Since $\beta$ is smaller but close to one, by (16) we can observe that, in dilute poroelastic media, the compressional wave has a velocity that is close but always higher than that one in the pure fluid phase.

Since (14) is a completely elastic wave equation, to take into account for the US energy dissipation, the highest order of approximation must be considered. In order to do that, we put (6) in (10) and with the help of (4) we obtain the wave equation

$$\nabla^2 \frac{R}{\beta} e_{\alpha\alpha} \cong \beta \,\partial^2 (\beta\rho_f e_{\alpha\alpha})/\partial t^2 + (1-\beta)(\rho_{11}/\beta f F_{(\omega)})\partial^3(\rho_f e_{\alpha\alpha})/\partial t^3 \qquad (17)$$

### *1. Phase velocity and attenuation of "fast" plane wave*

Equation (17) for plane waves $e_{\alpha\alpha} = C_I e^{-\alpha x} e^{i(kx-\omega t)}$ gives the characteristic equation

$$(k+i\alpha)^2 = \omega^2 \rho_f \left(\beta^3/R\right) \cdot \left[1 - i\omega(Q/R)\cdot\left(\rho_{11}/f F_{(v)}\right)\right] \qquad (18)$$

By solving $k$ and $\alpha$, and by using the purely elastic phase velocity $c_0^2 = R/\rho_f \beta^3$, the speed and the attenuation per cycle are obtained respectively

$$c^2 = c_0^2 \left(1 + \alpha^2/k^2\right) \qquad (19)$$

$$2\pi\alpha/k = -\pi(c/c_0)^2 \omega(1-\beta)\cdot\left(\rho_{11}/\beta f F_{(\omega)}\right) \qquad (20)$$

**B. Compressional fast wave in hydrogels**

When we introduce a new constituent such as the bounded water, we must pay attention to the definition of the fluid volume fraction of the hydrogel.

Since the bounded water phase is not a fluid phase, it must be subtracted by the water volume fraction "$\beta$" that is the total volume fraction of the water.

The effective free water volume fraction $\beta_{eff}$ can be obtained on the hypothesis that the number of bounded water molecules is proportional to the polymer-water contacts.

Since the probability of a polymer-polymer contact at very low polymer concentration (near $\beta \approx 1$, let's say for $1 - \Delta << \beta < 1$) is practically null, $\beta_{bw}$ is proportional to the polymer volume fraction (1- $\beta$) in that range.

When the polymer concentration is high (let's say for $0 < \beta << 1 - \Delta$)), due to the high probability of polymer-polymer contacts, the increase of polymer content does not cause a proportional increase of bounded water, so that $\beta_{bw}$ must approach a constant value $\phi$ as $\beta$ goes to zero.

If we approximate the approaching of the bounded water volume fraction to this constant value $\phi$ by means of an exponential law, we can write

$$\beta_{bw} = \phi \left(1 - \exp\left[-(1-\beta)/\Delta\right]\right) \tag{21}$$

where $0 < \Delta < 1$ and $0 < \phi < 1$ are empirical parameters to be deduced from the experimental data.

Therefore, the free water volume fraction reads

$$\beta_{eff} = \beta - \phi \left(1 - \exp\left[-(1-\beta)/\Delta\right]\right) \tag{22}$$

Moreover, given $\Delta << \varepsilon$, for $\beta \leq (1 - \varepsilon)$ it holds that: $\beta_{bw} \cong \phi$ and $\beta_{eff} \cong \beta - \phi$.

By introducing the free water volume fraction $\beta_{eff}$, the pure elastic acoustic fast wave (15) and its speed respectively read

$$\nabla^2 R e_{\alpha\alpha} \cong \beta_{\text{eff}}^{3} \partial^2 \left(\rho_f e_{\alpha\alpha}\right) \big/ \partial t^2 \qquad (23)$$

$$c_0^2 \cong \frac{c_f^2}{\left(\beta - \phi(1 - \exp[-(1-\beta)/\Delta])\right)^3} \qquad (24)$$

where $c_0 = (R/\rho_f \beta_{\text{eff}}^3)^{1/2}$ is the pure elastic longitudinal wave velocity, $c_f = (R/\rho_f)^{1/2}$ its velocity in the intermolecular fluid (free water), R is Biot's compressional modulus of the fluid[16,21] and $e_{\alpha\alpha}$ is the trace of the free water stress tensor.

Eq. (24) for $\Delta \ll \varepsilon$ and $\beta \leq (1 - \varepsilon)$ (i.e., $(1-\beta)/\Delta \gg 1$) leads to the relation

$$c_{0(\beta)}^2 \cong \frac{c_f^2}{(\beta - \phi)^3} \qquad (25)$$

Therefore, for $\varepsilon \ll 1$, the amount of bounded water $\phi$ can be evaluated by means of the best fitted value $\lim_{\beta \to 1} c_{0(\beta)}$ of the experimental data $c_{0(\beta_{(i)})}$ for $\beta_{(i)} \leq (1 - \varepsilon)$), following the equation

$$\phi = 1 - \left(\frac{c_f}{\lim_{\beta \to 1} c_{0(\beta)}}\right)^{2/3} \qquad (26)$$

It is noteworthy to note, when $\varepsilon \ll 1$, that the presence of the bounded water generates a velocity gap between the US velocity in pure water and that one extrapolated for hydrogels at $\beta = 1$ (a hydrogel made up by 100% of water).

*1. Bounded water-polymer network viscoelastic interaction*

The bounded water-polymer network viscoelastic interaction, can be introduced in the poroelastic equations by adding to Biot's viscous force (with $F_{(\omega)} = 1$)

$$\beta_e f \; \partial(e^*_{\alpha\alpha} - e_{\alpha\alpha})/\partial t \qquad (27)$$

the viscoelastic one due to the bounded water

$$\eta_{(\omega)} \partial(\varepsilon_{\alpha\alpha} - e^*_{\alpha\alpha})/\partial t + \kappa(\varepsilon_{\alpha\alpha} - e^*_{\alpha\alpha}) \qquad (28)$$

where $e^*_{\alpha\alpha}$ is the traces of the bounded water stress tensor.

Actually, introducing the bounded water stress tensor variable, the bi-phasic approach disembogues into a three-phasic one that is out of the purpose of this work. Actually, the bi-phasic model can be retained since bounded water volume fraction is very small. In this case, the biphasic mean can be conceived composed by the "polymer-bounded water aggregate" matrix plus the interstitial free water. In this way it is possible to introduce the following approximations:

  i. The polymer mass density can be assumed equal to that one of the solid aggregate.
  ii. The inertial effect of bounded water can be disregarded;
  iii. The trace of the strain tensor of the polymer $\varepsilon_{\alpha\alpha}$ approximates that one of the solid aggregate.
  iv. P,Q,R are the poroelastic constants of the new bi-phasic mean;

By means of the second hypothesis, the force between the free water and the bounded water can be equated to that one between the bounded water and the polymer matrix leading to the additional equation needed to solve the three stresses that reads

$$\beta_e f F_{(\omega)} \; \partial(e^*_{\alpha\alpha} - e_{\alpha\alpha})/\partial t = \eta_{(\omega)} \partial(\varepsilon_{\alpha\alpha} - e^*_{\alpha\alpha})/\partial t + \kappa(\varepsilon_{\alpha\alpha} - e^*_{\alpha\alpha}) \qquad (29)$$

Moreover, under the hypotheses (i, ii-iv) the system of poroelastic equations (1,2) reads

$$\nabla^2(P\varepsilon_{\alpha\alpha} + Qe_{\alpha\alpha}) = \partial^2(\rho_{11}\varepsilon_{\alpha\alpha} + \rho_{12}e_{\alpha\alpha})/\partial t^2 + \beta f \; \partial(e^*_{\alpha\alpha} - e_{\alpha\alpha})/\partial t \qquad (30)$$

$$\nabla^2(Q\varepsilon_{\alpha\alpha} + Re_{\alpha\alpha}) = \partial^2(\rho_{12}\varepsilon_{\alpha\alpha} + \rho_{22}e_{\alpha\alpha})/\partial t^2 - \beta f \; \partial(e^*_{\alpha\alpha} - e_{\alpha\alpha})/\partial t \qquad (31)$$

Given that in dilute bi-phasic means the compressional plane wave of interstitial fluid $e_{\alpha\alpha} = C_I e^{-\alpha x} e^{i(kx-\omega t)}$ induces a planar "slow" wave, as shown by (13) that gives $(\varepsilon_{\alpha\alpha} - e_{\alpha\alpha}) = -(i\omega Q \beta_e^2 / Rf) e_{\alpha\alpha}$; equation (29) in this case reads

$$\beta_e f \ \partial(e^*_{\alpha\alpha} - e_{\alpha\alpha})/\partial t = (\eta_{(\omega)} + j(\kappa/\omega)) \ \partial(\varepsilon_{\alpha\alpha} - e^*_{\alpha\alpha})/\partial t \qquad (32)$$

leading after simple manipulation to the relation

$$\partial(\varepsilon_{\alpha\alpha} - e_{\alpha\alpha})/\partial t = \{(\beta_e f + \eta_{(\omega)} + j(\kappa/\omega))/(\eta_{(\omega)} + j(\kappa/\omega))\} \partial(e^*_{\alpha\alpha} - e_{\alpha\alpha})/\partial t \qquad (33)$$

that can be recast as

$$\beta_e f \ \partial(e^*_{\alpha\alpha} - e_{\alpha\alpha})/\partial t = F^*_{(\omega)} \ \partial(\varepsilon_{\alpha\alpha} - e_{\alpha\alpha})/\partial t \qquad (34)$$

where the complex "friction" coefficient $F^*(\omega)$ reads

$$F^*(\omega) = [(\beta_e f)^{-1} + (\eta_{(\omega)} + j(\kappa/\omega))^{-1}]^{-1} \qquad (35)$$

Hence, by introducing (34) in (30,31), for plane waves, the dilute matrix equations (17) reads:

$$\nabla^2 \frac{R}{\beta} e_{\alpha\alpha} \cong \beta \partial^2 (\beta \rho_f e_{\alpha\alpha})/\partial t^2 + (1-\beta)(\rho_{11}/F^*_{(\omega)})\partial^3(\rho_f e_{\alpha\alpha})/\partial t^3 \qquad (36)$$

so that for the fast plane wave $e_{\alpha\alpha} = C_I e^{-\alpha x} e^{i(kx-\omega t)}$ it holds the characteristic equation

$$(k + i\alpha)^2 = \omega^2 \rho_f (\beta_e^3/R) \cdot \left[1 - i\omega(1-\beta_e)\left(\rho_{11}/F^*_{(\omega)}\right)\right] \qquad (37)$$

that solved in $\alpha$ and $c$ gives

$$c^2 = c_0^2 \left(1 + \alpha^2/k^2\right) \Big/ \left[1 + (1-\beta_e) \cdot \omega \cdot \rho_{11} \operatorname{Im}\left\{F^{*-1}_{(\omega)}\right\}/\beta_e^2\right] \qquad (38)$$

$$\alpha/k = -1/2 (c/c_0)^2 \omega(1-\beta_e)\rho_{11} \operatorname{Re}\left\{F^{*-1}_{(\omega)}\right\} \qquad (39)$$

where

$$\operatorname{Re}\left\{F^{*}_{(\omega)}{}^{-1}\right\} = \left[\eta_{(\omega)}\left(1+\eta_{(\omega)}/\beta_e f\right)+\left(\kappa^2/\beta_e f \omega^2\right)\right] \Big/ \cdot \left(\eta_{(\omega)}{}^2 + \left(\kappa/\omega\right)^2\right) \quad (40)$$

$$\operatorname{Im}\left\{F^{*}_{(\omega)}{}^{-1}\right\} = -\left(\kappa/\omega\right) \Big/ \cdot \left(\eta_{(\omega)}{}^2 + \left(\kappa/\omega\right)^2\right) \quad (41)$$

In the following we model the case when the bounded water-network interaction is prevalently viscous such as

$$\eta > (\kappa \rho_{bw})^{\frac{1}{2}} \quad (42)$$

where $\rho_{bw}$ is the mass density of the bounded water that we assume close to that of free water $\rho_f$.

At high frequencies such $\omega > \omega_g = 2\pi\, \eta\, /\, \rho_f > (\kappa\, /\rho_{bw})^{\frac{1}{2}}$ (far away from bounded water resonance) hence, it holds

$$\kappa\, /\omega\, \eta \ll \kappa\, /\omega_g\, \eta < 1 \quad (43)$$

Moreover, in order that the crossover frequency $v_g = \eta\, /\, \rho_f$ be much smaller than the unattainable Biot's one $v_c = f/\rho_f$, in the high frequency limit, it must hold

$$\lim_{\omega \to \infty} \eta_{(\omega)} \ll f \quad (44)$$

If the high frequency behavior of the bounded water viscosity $\eta_{(\omega)}$ is modeled by a dimensionless variable as

$$\eta_{(\omega)} = \eta_0\, (\omega_g\, /\omega)^{\delta} \quad (45)$$

it follows that $\omega_g = 2\pi \nu_g = 2\pi \eta_{(\omega=\omega_g)} / \rho_f = 2\pi \eta_0 / \rho_f$, and that conditions (43) and (44) hold contemporarily by requiring $0 < \delta \leq 1$. In such a case, by introducing them in (40,41) it follows that

$$\lim_{\omega/\omega_g \gg 1} \text{Re}\left\{F^*_{(\omega)}{}^{-1}\right\} \cong (\kappa/\omega\eta)2 + \left(1 + \eta_{(\omega)}/\beta_e f\right) / \eta_{(\omega)} \cong 1/\eta_{(\omega)} \qquad (46)$$

$$\lim_{\omega/\omega_g \gg 1} \text{Im}\left\{F^*_{(\omega)}{}^{-1}\right\} \cong 0 \qquad (47)$$

That in a dilute polymer hydrogel ($\beta \approx 1$ and $\rho_{11} \ll 1$) leads to

$$c^2 = c_0^2 \left(1 + \alpha^2/k^2\right) \cong c_0^2 \qquad (48)$$

being typically $(\alpha/k)^2$ very small (of order of $10^{-3}$ in our experiments); and to the specific US attenuation per cycle

$$2\pi\alpha/k \approx -\pi\left(c/c_0\right)^2 \left(\omega/\omega_g\right)^{1+\delta} \left(1 - \beta + \phi(1 - \exp[-(1-\beta)/\Delta])\right) \cdot \sigma_{pf} \qquad (49)$$

where $\sigma_{pf} = \rho_{11}/\rho_f$.

For hydrogels with small $\Delta$, and for $(1 - \beta)/\Delta \gg 1$, finally, (49) leads to

$$2\pi\alpha/k \approx -\pi\left(c/c_0\right)^2 \left(\omega/\omega_g\right)^{1+\delta} \left(1 - \beta + \phi\right) \cdot \sigma_{pf} \qquad (50)$$

From Fig. 1 it can be observed that the bounded water raises the absorption straight line, while $\Delta$ smoothes the curve in the interval $1 - \Delta > \beta > 1$

From Fig. 2 it can be seen that:

i. the bounded water leads to an upward shift of the wave speed
ii. that $\Delta$ smoothes the curve upwards in the region $1 - \Delta < \beta < 1$

iii. when ε << 1 there is a speed jump at $\beta \cong 1$.

**D. High polymer concentration**

For the equations (30,31) the characteristic equation for slow and fast plane waves turns out to be formally the same as that one given by Biot[15] where the friction term $\beta fF_{(\omega)}$ is substituted by $F^*(\omega)$ (see (34)) that at high frequency is given by (46,47).

Nevertheless, it must be observed that the Biot's model[15] assumes that the bi-phasic material parameters are constants (or at least very smoothly varying) while in hydrogels they actually change very much at low water volume fraction[27]. Moreover, since our intent here is limited to understand the range of validity of the dilute matrix approximation and eventually to improve the experimental fitting of data at intermediate water volume fraction values (but not to validate the model), we opt for a more direct semi-empirical approach.

In order to do that, we observe that for vanishing $\beta = 0$ values, the speed of propagation of the US is finite and equal to that of polymeric solid $c_S$, while the law (24) diverges to infinity at $\beta = \phi$ and gives negative values for $\beta = 0$.

The contribution that cancels this divergence can be simply written as a term (let's call it $G(\beta)$) added to the denominator of (24) to read:

$$c^2 = \frac{c_f^2}{\left(\beta - \phi(1 - \exp[-(1-\beta)/\Delta])\right)^3 + G_{(\beta)}} \tag{51}$$

where the only conditions to require to $G(\beta)$ are: $\lim_{\beta \to 1} G_{(\beta)} = 0$ and $\lim_{\beta \to 0} G_{(\beta)} > 0$.

for which it is appropriate to use the series approximation: $G(\beta) = \gamma(1-\beta) + \chi(1-\beta)^2$ with $\gamma + \chi > 0$, that in (29) leads to

$$c^2 = \frac{c_f^2}{\left(\beta - \phi(1 - \exp[-(1-\beta)/\Delta])\right)^3 + \gamma(1-\beta) + \chi(1-\beta)^2} \tag{52}$$

and that for $(1-\beta)/\Delta \gg 1$ gives

$$c^2 = \frac{c_f^2}{(\beta - \phi)^3 + \gamma(1-\beta) + \chi(1-\beta)^2} \quad (53)$$

From Fig. 3 we can see the change of the US wave speed following the introduction of the polynomial expression $\mathcal{G}(\beta)$: the polymer network cancels the divergence of US speed at $(\beta - \phi) = 0$ and lowers it at the intermediate values of $\beta$ in a progressive manner.

Finally, for $\beta$ close to 1 we observe that the increase of the bounded water in (53) increments the US speed through the term $(\beta - \phi)^3$ while the elasticity of the polymer network lowers it mainly through the vanishing term $\gamma(1-\beta)$ (if $\gamma > 0$).

**III. EXPERIMENTAL**

**A. Materials and methods**

Gel samples were prepared in parallelepipeds of $2 \times 2 \times 1 \text{cm}^3$ using polyvinylalcohol (PVA) with a degree of hydrolysis of 99+ ‰ and an average molecular weight of 115.000 ± 30.000 (Aldrich, Milan, Italy), dissolved into de-ionized water in a concentration of 10% by weight. The homogeneous solution was refrigerated at - 80 °C starting from room temperature 23 ± 1 °C. The freezing-thawing procedure was repeated 2, 3, 4 and 6 times. The samples were left to equilibrate in de-ionized water for 72 hours.

The experimental tests were carried out at different hydration conditions down to a minimum of about 50%, taking care that the drying was gradual and homogenous.

The reversibility of the de-hydration treatment was checked at the end after the drying steps. The ultrasonic pulses were generated by the Panametrics® Pulser model 5052PR coupled with a PVDF piezoelectric transducer obtained in our laboratory following the Naganishi e Ohigashi procedure[28].

The Pulser voltage spike of -270 V induces an ultrasonic pulse of 250KHz of frequency through the transducer.

The distance between the transducer and the reflecting iron layer behind the samples was measured with an accuracy of ± 0.01 cm and maintained constant through the whole experiment.

Echo Signal registration and conditioning data were collected with a routine and carried out with the LabView™ software on a computer through a National Instruments® DAQ device.

Finally, the samples were totally dehydrated in an oven at 40°C with desiccant silica gels, to measure their polymer content and the US speed in the dry solid.

The US absorption coefficient "$\alpha$" was deduced by using the mathematical relation $\alpha = \frac{1}{2d} \ln \frac{A_0}{A_{(2d)}}$, where $A_0$ and $A_{(2d)}$ represent both the initial and final wave amplitude, respectively, and where $d$ is the sample thickness.

The water volume fraction of the hydrogel samples $\beta = V_w / (V_w + V_p)$, where $V_w$ and $V_p$ are the volume of water and polymer respectively, was obtained by means of the respective weight fractions $P_w$ and $P_p$ such as $\beta \approx P_w / (P_w + P_p)$ since the water and PVA specific densities are very close each other.

After the samples were left to fully hydrate themselves in distilled water, the fractional water volume was measured to be $\beta = 0{,}917 \pm 0.001$ for all the samples.

The fitting of the experimental results were carried out by means of a multiple parameter best fit utilizing an appropriate routine in MATLAB® 7.0.

**B. Ultrasound phase velocity**

Experimental data of US wave speed with the best-fitted curve are shown in Fig. 4 for the hydrogel sample submitted to two cycles of cross-linking. The most probable values of the model constants for all samples are shown in table I.

From Fig. 4 we can see that at intermediate values of $\beta$, the US wave speed progressively departs by the model dependence $(\beta - \phi)^{-3/2}$ towards the value of the dry solid polymer in agreement with the theoretical behavior (52) shown in Fig.3. The measurements show that the effect of the polymer matrix starts to influence the US speed of propagation for values of $\beta$ as high as $0.8 \div 0.9$.

For $\beta$ close to zero, the US speed of propagation converges to the speed of the dry polymer which rises when the number of cross-linking cycles increases because of the greater polymer stiffness as shown in table I.

In Fig. 5 the US speed is extrapolated for $\beta$ equal to one by means of a best fit procedure. The data show that the US phase velocity is sensibly higher than that of pure water (that in our experimental condition has been measured to be 1483 m/s). In Fig.6 the bounded water volume fraction "$\phi$"is shown as a function of the number of cross-linking cycles of the sample. The evaluation of "$\phi$" is obtained by introducing in (26), being $\Delta \cong 0$, the limiting velocity values of the fits of Fig.5 and given in table I. The bounded water volume fraction $\phi$ ranges from 2 % in the PVA samples with six cross-linking cycles, to 13 % for the PVA samples with two cross-linking cycles.

As shown in table I, we can observe that the lower the gap speed at $\beta = 1$, the bigger the number of the cross-linking cycles and the lower the measured bounded water fraction $\phi$ present in the hydrogel. This agrees with the characteristics of PVA hydrogel synthesized by means of freezing-thawing cycles where the number of polymer-polymer contacts (cross-links) grows as the number of freezing-thawing cycles increases[29]. Moreover, since lower values of $\phi$ in samples with higher cross-linking correspond to lower values of the US speed, this behavior cannot be ascribed to the elasticity of the polymer network that leads to a variation of the opposite sign.

In fact, as shown in table I, both $\gamma$ and $\gamma + \chi$ (that account for the effect of the polymer network elasticity on the US speed, ($\gamma$ contributes prevailing in the region of high $\beta \approx 1$, while $\gamma + \chi$ in the region of small $\beta \approx 0$)), decrease with the increase of network cross-linking and by looking at (53) it comes clearly out that in such a case they would cause an increase of the US speed.

Moreover, since both $\gamma$ and $\gamma + \chi$ are positive, for the intermediate values of $\beta$, the effect of the polymer network is to decrease the US speed and to eliminate the US speed divergence due to the dilute matrix approximation for $\beta$ approaching zero.

From experimental data we can also see that the value of the parameter $\Delta$ obtained from the last square fit procedure is equal to 0.0006 (practically null) for all PVA gel samples. This fact says that the bounded water inside the PVA hydrogel samples remains practically constant as a function of $\beta$ and that only free water is subtracted when they are mildly dried.

**C. Ultrasound wave attenuation**

Some of the experimental data as well as the fits of the theoretical expression (49) are shown in Fig. 7. For the attenuation measurements, as well as for the velocity ones, the value of the fitted parameter $\Delta$ is resulted practically null.

A linear decrease of the US wave attenuation "$\alpha$" is measured as a function of $\beta$ for all samples with a mean value $\partial\alpha/\partial\beta = -2.4 \pm 0.4$ cm$^{-1}$.

The pure water attenuation has also been measured to be 0.022 cm$^{-1}$ and found to be much smaller than that one of the gel in the order of 1 cm$^{-1}$.

**IV. DISCUSSION**

Even if the equation that well describes the US propagation in natural hydrogels and soft tissues is available[30-31], it is indeed like a box where the parameters can be empirically adjusted to describe the propagation of acoustic waves.

Yang and Cleveland[32] show that it is possible to reproduce the frequency dependence of the absorption with an exponent equal to 1.1 by supposing two juxtaposed relaxing processes that may or may not be real. In this way some information about the material mean (e.g., a natural gel or a tissue) is lost from the US behavior.

The model presented here shows how the behavior of the US propagation is linked to the structure of the mean.

More recently Kowalski[33] developed a theory for the propagation of US in a dilute suspension. Even if conceptually different from a polymeric gel, a fluid suspension can be assimilated to a bi-phasic mean with a solid matrix (with an its own elasticity and permeability) and a fluid that permeates the interstices. Obviously, the Kowalski model is expressed in terms of the viscoelastic parameters of the suspension that do not correspond tout court to those of the biphasic approach, except for the fluid fraction volume $\beta$.

The theoretical dependence of the US velocity as a function of the water volume fraction found by Kowalski is of the same type obtained in the present paper with the same good experimental agreement. The main difference is that the suspension does not have bounded water (i.e., $\phi = 0$) so that at $\beta = 1$ the US speed converges to that of the pure fluid.

Hence, the following features of the US propagation in hydrogels can be ascribed to their structure:

i. The speed of propagation of US in highly hydrated gels is always a little bit higher than that of water (1480 m/s) since in (26) $\beta < 1$ and $\phi > 0$.

ii. The US specific attenuation "$2\pi\,\alpha/k$" in natural hydrogels can follow a non-integer frequency law $\nu^{(1+\delta)}$ with $0 < \delta \leq 1$ where $\delta$ is the exponent of the polymer-bounded water viscous interaction.

This outcome well agrees with the attenuation data of US in natural gels where $\delta$ is typically about $0.25 \div 0.50$ while the value for pure water is 1.

It must be also noted that the possibility of having a model in which the correspondence between the mean structure and the US behavior is disclosed can help the development of tissue-mimicking US phantoms.

Moreover, by modeling biological cells as poroelastic spheres, endowed by internal and superficial elasticity as well as permeability, dispersed in a hydrogel environment, a bi-phasic model of soft natural tissue can be obtained. In this case, the possibility to describe the US propagation in terms of collective cells characteristics opens up new perspectives on US non-invasive tissue physiology evaluation since the permeability and elasticity of the cell structures depend strongly on their state and functionality.

## V. CONCLUSIONS

The continuum model for US acoustic longitudinal waves in hydrogels proposed in the present paper satisfactory agrees with the experimental results.

The experimental measurements reveal that when the hydrogel water volume fraction $\beta$ is close to one (100 % water) a gap exists between the measured US speed of propagation in the gel samples and that of pure water.

The proposed poroelastic model shows that the speed gap, at $\beta = 1$, is due to the presence of bounded water around the hydrogel polymer matrix: the bigger the bounded water concentration, the higher the US speed gap at $\beta = 1$.

The experiments show that gel samples with a lower degree of cross-linking (and hence, with higher bounded water volume fraction) have a higher US speed gap at $\beta = 1$. The evaluated bounded water fraction $\phi$ ranges from 2% in the PVA samples with six cross-linking cycles, to 13 % for the PVA samples with two cross-linking cycles.

In the range of experimental conditions ($\beta > 0.4$), the bounded water volume fraction $\phi$ inside each type of PVA hydrogel does not depend by the free water content. As the gel water fraction $\beta$ is lowered towards zero, the US propagation speed is more and more influenced by the matrix elasticity and increases in samples with a higher network cross-linking in agreement with the classical rheological law.

The experimental data also show that the US attenuation in hydrogels decreases with the increase of the water volume fraction $\beta$ in a linear way in agreement with the theoretical forecast.

TABLE I

Most probable values of the model parameters $\phi$, $\Delta$, $\gamma$, $\chi$, and $\gamma + \chi$, obtained by the best fit of the experimental data by means of the expression (31) as a function of the number of cross-linking cycles (first column).

| Number of cross-linking cycles | $\phi$ | $\Delta$ | $\gamma$ | $\chi$ | $\gamma + \chi$ | $\lim_{\beta \to 1} c$ <br> m s$^{-1}$ | $c_s$ <br> m s$^{-1}$ |
|---|---|---|---|---|---|---|---|
| 2 | 0.131 | 0.0006 | 0.742 | -0.453 | 0.289 | 1830 | 2755 |
| 3 | 0.075 | 0.0006 | 0.570 | -0.363 | 0.206 | 1667 | 3266 |
| 4 | 0.063 | 0.0006 | 0.445 | -0.247 | 0.197 | 1636 | 3339 |
| 6 | 0.020 | 0.0006 | 0.307 | -0.109 | 0.198 | 1528 | 3340 |

**FIGURES CAPTIONS**

Figure 1: Theoretical behavior of acoustic US fast wave absorption in a hydrogel given by Eq. (31) as a function of the water volume fraction for various values of the parameters: $\phi = 0$, $\Delta = 0$, (full line); $\phi = 0.2$, $\Delta = 0.0006$, (dashed line); $\phi = 0.2$, $\Delta = 0.1$, (dashed-dotted line).

Figure 2: Theoretical behavior of the acoustic speed of propagation of US fast wave in a hydrogel given by Eq. (4) as a function of the water volume fraction for various values of the parameters: $\phi = 0$, $\Delta = 0$, (full line); $\phi = 0.1$, $\Delta = 0$, (dashed line); $\phi = 0.1$, $\Delta = 0.1$, (dashed-dotted line).

Figure 3: Theoretical behavior of the acoustic speed of propagation of US fast wave in a hydrogel given by Eq. (31) as a function of the water volume fraction for various values of the parameters: $\phi = 0.1$, $\Delta = 0$, $\gamma = 0.7$, $\chi = -0.4$, (dashed line); $\phi = 0.1$, $\Delta = 0.1$, $\gamma = 0.7$, $\chi = -0.4$, (full line).

Figure 4: Experimental acoustic US speed of propagation of fast wave in PVA hydrogels with best fit (full line) as a function of the water volume fraction for the hydrogel sample with two cycles of cross-linking.

Figure 5: Experimental acoustic US speed of propagation of fast wave in PVA hydrogels as a function of the water volume fraction at various degrees of polymer matrix cross-linking: (□) two cycles of cross-linking with best fit (full line); (○) three cycles of cross-linking with best fit (dotted line); (✘) four cycles of cross-linking with best fit (dashed-dotted line); (▽) six cycles of cross-linking with best fit (dashed line).

Figure 6: Estimated bounded water fraction in PVA hydrogel as a function of the number of cross-linking cycles.

Figure 7: Experimental acoustic US fast wave attenuation in PVA hydrogels as a function of the water volume fraction at various degrees of polymer matrix cross-linking: (▲) four cycles of cross-linking with best fit (dotted line); (+) six cycles of cross-linking with best fit (dashed-dotted line).

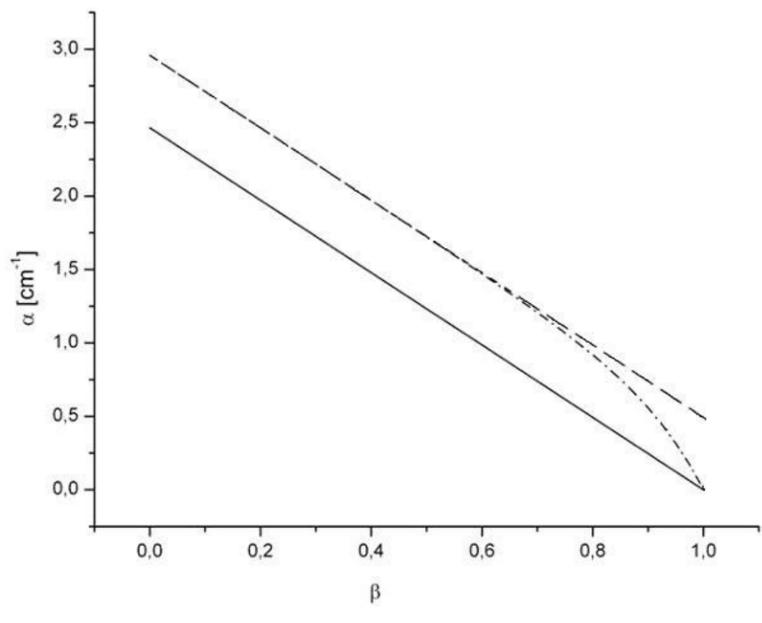

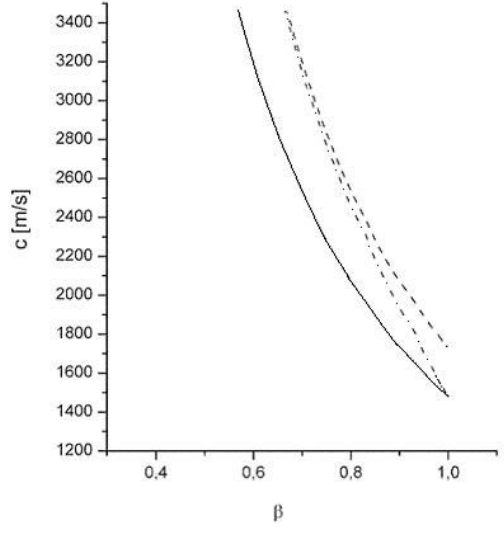

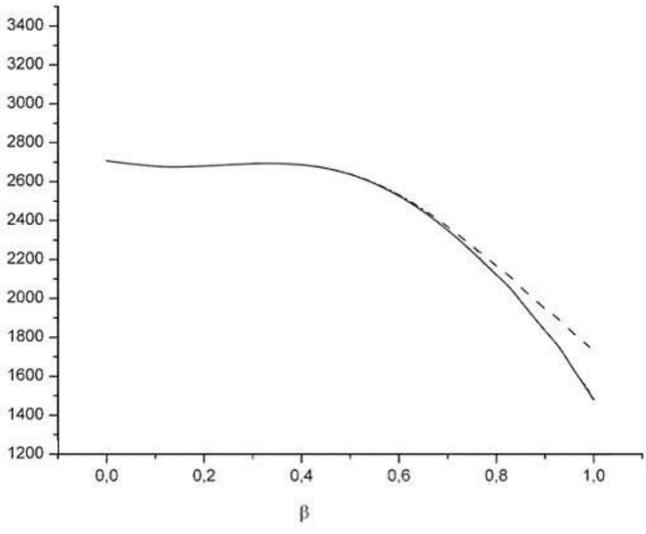

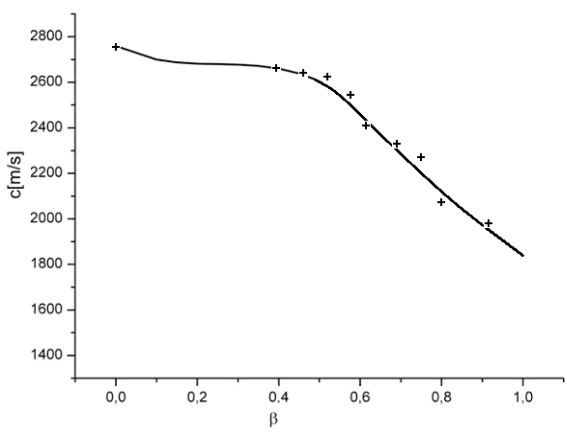

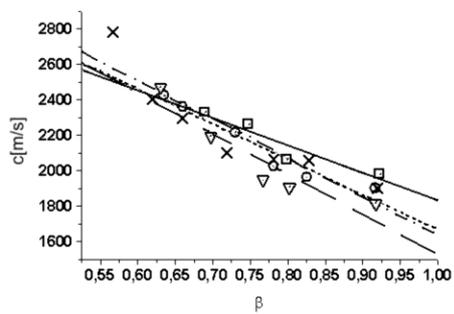

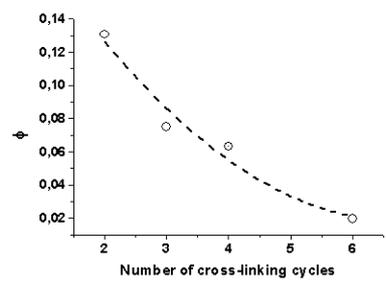

Number of cross-linking cycles

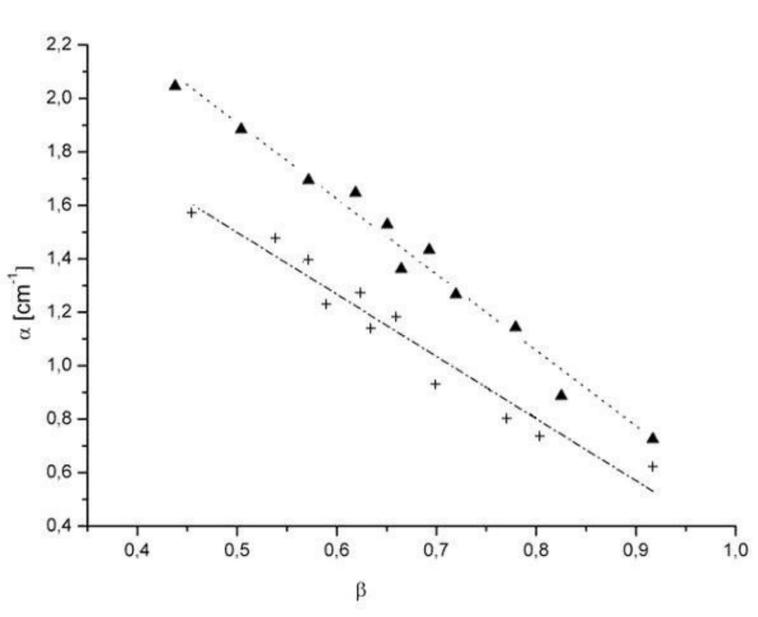